\documentclass[twocolumn,showpacs,preprintnumbers,amsmath,amssymb]{revtex4}
\usepackage{graphicx}
\input{epsf}

\newcommand{\sig}{\mbox{\boldmath{$\sigma$}}}

\usepackage{graphicx}

\input{epsf}


\begin{document}
\title{Intervalley scattering by  charged impurities in graphene}

\author{L.S. Braginsky and M.V. Entin}
\email{brag@isp.nsc.ru, entin@isp.nsc.ru} \affiliation{Institute of Semiconductor
Physics, Siberian Branch, Russian Academy of Sciences,
Novosibirsk, 630090 Russia\\Novosibirsk
State University, Novosibirsk, 630090 Russia}

\date{\today}

\begin{abstract}
Intervalley  charged-impurity scattering processes are examined. It is found that the scattering probability is enhanced due to the Coulomb interaction with the impurity by the Sommerfield factor $F_Z\propto \epsilon^{2\sqrt{1-4g^2}-2}$, where $\epsilon$ is the electron energy and $g$ is the dimensionless constant of the Coulomb interaction.
\end{abstract}
\maketitle
\subsection*{Introduction}
The presence of  multiple valleys  is an ordinary situation in semiconductors; the intervalley scattering, e.g. in Si and Ge, has been studied since 1950th. Large distance between the valleys makes the transitions between them  difficult, as compared with the intravalley processes. Therefore,  the valley population becomes a well-conserving quantity that determines different properties of such semiconductors. By analogy with the ordinary spin, the valley number can be treated as a new quantum number  ''pseudospin'',  which
determines the long-living electron states in semiconductors. The processes caused by  different population of the equivalent valleys, in particular, surface photocurrent and polarized photoluminescence were studied   long ago (see, e.g., \cite{efanent0,efanent} and references therein).

The processes involving a different valley population gave birth of the promising
new electronic device applications called valleytronics  ~\cite{TarIvch,Rycerz,Karch}.
The valley-polarized current can be emerged in the graphene point contact with zigzag edges,~\cite{Rycerz}, the graphene layer with broken inversion symmetry,~\cite{Xiao} or under illumination of the circularly  polarized light~\cite{Oka}.

The study of the valley dynamics  attracted  attention to the intervalley relaxation that controls the valleys population.
Recent interest to graphene has been mainly focused on its conic electron spectrum. However, the presence of two different valleys has been remained out of interest for a long time.  Meanwhile, just low DOS near the cone point supposedly suppresses the intervalley transitions and makes the valley population long-living.

The non-equilibrium between two graphene valleys means the violation of both spatial and time reversibility.
As far as the spatial irreversibility determines the valley photocurrents \cite{TarGolEntMag}, the time reversibility is
responsible for  weak localization \cite{McCann}, thereby the valley relaxation time is an  important electronic parameter of graphene.

The valley relaxation is determined by the processes with a large momentum transfer, and, therefore, its scattering length is of the order of the lattice constant.
At the same time, the Coulomb impurity determines the interaction on the large distances.
Consequently, the probability of the electron penetration to the short-scale impurity core, where it experiences
intervalley scattering,
is determined by the large-scale wave function behavior   and strongly depends on the electron energy.
In particular, the Coulomb attraction or repulsion to impurity should essentially affect this process.

The purpose of the present paper is to study the intervalley charge-impurity scattering in the monolayer graphene. We consider the problem in the envelope-function approximation. The solution of the impurity scattering problem will be found in the
Born approximation.  Then, the Coulomb solution will be applied to the renormalization of the Born short-range scattering result.

\subsection*{Problem Formulation}
We use the two-atom basis of  graphene $|a>$ and $|b>$. The tight-binding Hamiltonian for the ideal
graphene in the momentum representation reads

\begin{equation}\label{Hamiltonian}
\hat{H}_{00}= \left(
\begin{array}{cc}
0 & \Omega_{\bf p} \\
\Omega_{\bf p}^* & 0
\end{array}
\right) \:.
\end{equation}
%\begin{equation}\label{Omega}
Here $\Omega_{\bf p} =
t(e^{ip_ya}+2e^{-ip_ya/2}\cos(p_xa\sqrt{3}/2)),
$
%\end{equation}
$a=0.246$ nm is the lattice constant, and $t$ is the tunnel amplitude. The energy is counted from the  permitted band center.

The long-range Coulomb interaction
with an impurity $\hat{V}$ should be situated on the diagonal of the matrix, while the sort-range interaction with the impurity core  gives  the off-diagonal operator $\hat{U}$:
\begin{equation}\label{Ham}
\hat{H} =\hat{H}_{00} +\hat{V} +\hat{U}.\end{equation}
Here $\hat{V}$ and $\hat{U}$ represent the long- and short-range interactions with the impurity. In the case of the Coulomb impurity in the envelope-function representation  $V(r)=e^2/\chi r$, where $\chi$ is a half-sum of the dielectric  constants of surrounding media, $\bf r$ is a 2D radius-vector in the graphene plane. The short-range part $\hat{U}$ of the interaction acts over the atomic distance at the impurity. It is specific for the
type of impurity.

Consider now the  states of free electrons. Near the conic points $\bf p=\pm K$, ${\bf K}=(2\pi/a\sqrt{3},0)$ the Hamiltonian $H_{00}$ can be
transformed to Eq.~(\ref{Hamiltonian}) with $\Omega_{\bm
p}^{(\nu)}=s(-\nu k_x+ik_y)$, where $s=3ta/2$, $\bf p=\pm K+k$, $\nu=\pm$.
The corresponding wave functions near the point $\bf K$ can be written as $(1,-\mbox{sign}(\epsilon)e^{i\varphi_{\bf k}})e^{i ({\bf K}+{\bf k}) \bf
r}/\sqrt{{\cal A}}$, where $e^{i\varphi_{\bf k}}=(k_x+ik_y)/k$, ${\cal A}$ is the patten area, the wave functions are normalized to the full surface. Below we assume $\epsilon>0$ corresponding to the case of electrons.

In view of  $\pm \bf K$ states, the Hamiltonian $H_{00}$ is splitted into two independent
Hamiltonians referred to the points $\pm K$:

\begin{equation}\label{Ham}
\hat{H}_0 = \left(
\begin{array}{cccc}
0 & -k_x+ik_y&0&0 \\
-k_x-ik_y & 0&0&0\\
0&0&0&k_x+ik_y\\
0&0&k_x-ik_y&0
\end{array}
\right) \:.
\end{equation}
The elements of the wave function (column of four terms) are  $|a,{\bf K}>$, $|b,{\bf
K}>$, $|a,{-\bf K}>$, $|b,-{\bf K}>$, respectively.

In the envelope-function approximation, the coordinate representation of the short-range interaction potential can be expressed as
\begin{equation}\label{UU}\hat{U}=\tilde{U}\delta({\bf r}),~~~\tilde{U}=(U_0+{\bf S}\sig).\end{equation}
In the
tight-binding model, the components $U_0=S_0(\epsilon_A+\epsilon_B)/2$, $S_z=S_0(\epsilon_A-\epsilon_B)/2$, were
$S_0=a^2/2$ is the graphene unite cell area, are determined by the  levels of $A$ ($\epsilon_A$) and $B$ ($\epsilon_B$) atoms of the cell in the origin,
while the components $S_x$ and $S_y$ are determined by the perturbation of the amplitude of transition between these atoms.
  The
amplitude of transition between these atoms   is mapped onto the vector ${\bf S}$ as $(S_x+iS_y)=tS_0$. In principle, the Hamiltonian $\hat{U}$ describes both the monomer and dimer impurities. Below we deal with the case of a single impurity at the A site with the perturbation of energy level $\delta \epsilon_A=2U_0, S_z=U_0$   without transition amplitude ($S_x=S_y=0$) perturbation.

In the envelope-function
approximation, the long-range Coulomb interaction mixes the states
within one cone.  It is located on the diagonal in the 4$\times$4 form of the Hamiltonian. Keeping in mind the divergency of the final result at the small distances; later on, we should
cut
off this divergency at the  lattice constant $a$.

The short-range Hamiltonian of interaction contains the matrix elements between the states $|{\pm\bf K,a}\rangle$ and $|{\pm\bf K,b}\rangle$.
The blocks in the left-up and right-down from the diagonal yield the intravalley mixing, while the blocks in the right-up and left-down from the diagonal relate to the intervalley matrix elements of the impurity potential.
  Although, these blocks are identical  in the model Eq.(\ref{UU}), ${\bf K}\to -{\bf K}$  blocks have, generally speaking, a lower order of magnitude.
  Roughly,
these elements are the Fourier harmonics of Coulomb potential
at the momentum $K$.

\subsection*{Short-range potential}
\begin{figure}[h]
\label{fig1}
\centerline{\epsfxsize=4cm\epsfbox{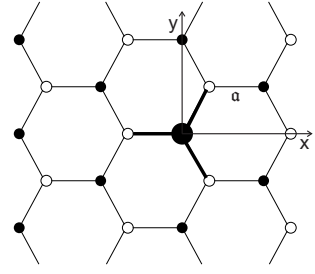}}

\caption{Graphene with substitution impurity (large circle), which energy  differs from that of the host atoms. }
\end{figure}
In this case $\hat{V}=0$.
The scattering amplitude is described by the t-matrix satisfying
the equation
\begin{equation}\label{short}
    t=U+UGt
\end{equation}
with the formal solution
\begin{equation}\label{short-solv2}
  t=T\delta(r),~~~  T=(1-\tilde{U}R)^{-1}\tilde{U},
\end{equation}
where $R_{\nu,\mu}$  is the projection of the Green function  onto the
origin lattice cell (e.g., (00)) populated by the impurity atom. 

The scattering probability is $2\pi|\langle\nu,{\bf k}|t|\nu',{\bf
k'}\rangle|^2\delta(\epsilon_{\nu,\bf k}-\epsilon_{\nu',\bf k'})$.

Eq. (\ref{short-solv2}) gives the symbolic solution of the short-gange scattering problem. Let us apply it to the Hamiltonian (\ref{UU}) with use of $R$
\begin{equation}\label{R}
   R=\int \frac{d^2p}{4\pi^2}\frac{1}{\epsilon^2-|\Omega_{\bf p}|^2}\left(
                                \begin{array}{cc}
                                  \epsilon & \Omega_{\bf p} \\
                                  \Omega_{\bf p}^* & \epsilon \\
                                \end{array}
                              \right),
\end{equation}
where the integration runs over the Brillouin zone. This integration gives a finite result even if $\epsilon\to 0$:
\begin{equation}\label{R0}
   R_0=\int \frac{d^2p}{4\pi^2}\left(
                                \begin{array}{cc}
                                 0 & 1/\Omega_{\bf p}^* \\
                                  1/\Omega_{\bf p}& 0\\
                                \end{array}
                              \right)+O(\epsilon\log(\epsilon a/s)).
\end{equation}

The amplitudes of the intra-  and inter-valley transitions in the Born approximation ($\hat{U}\to 0$) are
\begin{eqnarray}&&\nonumber A_{\bf K, k;K,k'}=\Big(U_0+S_z-(S_x+iS_y)e^{i\phi_{\bf k}}+\\&&(U_0-S_z)e^{i(\phi_{\bf k}- \phi_{\bf k'})}-(S_x-iS_y)e^{-i\phi_{\bf k'}}\Big)/{\cal A}\nonumber
\\\nonumber &&A_{\bf K, k;-K,k'}
    =\Big(U_0+S_z-(S_x+iS_y)e^{i\phi_{\bf k}}+\\&&(S_z-U_0)e^{i(\phi_{\bf k}+ \phi_{\bf k'})}+(S_x-iS_y)e^{i\phi_{\bf k'}}\Big)/{\cal A}
\end{eqnarray}
It should be emphasized that the transition probability has the essential angular dependence on the angles $\phi_{\bf k}$ and $\phi_{\bf k'}$. Besides, this dependence concerns not only the relative angle $\phi_{\bf k'}-\phi_{\bf k}$, but also the absolute angles. This dependence originates from the degeneracy of the states near the cone points and possible asymmetry of the defect. Note that such a
dependence is absent for the $\delta$-potential in the envelope-function approximation.
In the specific case of the monomer impurity, $A_{\bf K, k;-K,k'}=A_{\bf K, k;K,k'}=U_0/{\cal A}$.

If $\tilde{U}\to\infty$, $T\to 1/R$ and ceases to depend on $\tilde{U}$.

\subsection*{Electron states in the Coulomb potential}

The long-range Coulomb scattering does not change the valley. To
find the transition amplitude, we should use the intervalley block of
Hamiltonian $\hat{U}$. The Coulomb interaction in the ''final''
state corrects the amplitude. The Coulomb corrections to the wave
function are formed at the distances much exceed the
lattice constant. In that case $\hat{U}$  should be
multiplied by the limit of the Coulomb wave function at a low
distance from the impurity. This limit is determined by the
zero-momentum projection component of the wave function. The
Coulomb wave function should be matched with the free solution of
the equation without any potential.

The  equation with long-range Coulomb potential for two-component envelope wave function $(\phi,\chi)$ in the polar coordinates $(r,\varphi)$ reads
 \[\Bigg(\begin{array}{cc}
\epsilon-g/r&e^{-i\varphi}(i\partial_r+\frac{1}{r}\partial_\varphi)\\e^{i\varphi}
(i\partial_r-\frac{1}{r}\partial_\varphi)&\epsilon-g/r\end{array}\Bigg)
\Bigg(\begin{array}{c}\phi\\\chi\end{array}\Bigg)=0.\]
We search for
the solution of this equation with the substitution
\[\Bigg(\begin{array}{c}\phi\\\chi\end{array}\Bigg)=r^{1/2}\Bigg(\begin{array}{c}e^{i(M-1/2)\varphi}\phi_M\\
ie^{i(M+1/2)\varphi}\chi_M\end{array}\Bigg).\] Here
$M=m+1/2$, and $m$ is an integer. Then
\begin{eqnarray}
\label{1111}\nonumber(r\phi_M)'+(\epsilon r-g)
\chi_M-M\phi_M=0,\\(r\chi_M)'-(\epsilon r-g)
\phi_M+M\chi_M=0
\end{eqnarray}

The wave
function diverges at  small distances and has a divergent phase
(''falling down the center''). The integral of the electron density
converges, while the potential and the kinetic energies diverge. This
divergence is connected with  falling down the center. In fact, the conic approximation fails
at small distances from the center.
The problem can be resolved by introduction of a short-range
cutoff.

Eq.(\ref{1111}) corresponds to the Eq.(35.5) with solution Eq.(36.11) from
\cite{berestlandau} at $m=0$. Using these equations, we have the finite at $r=0$
solution
\begin{eqnarray}\label{a}
   && \Xi_m=\begin{array}{c}
      \phi_M \\
       \chi_M\\
    \end{array}\Big\}=(2\varepsilon r)^{\gamma-1}\nonumber\\
    &&\times\begin{array}{c}
      {\rm Im }\\
      {\rm Re} \\
    \end{array}
    \left\{
    e^{i(\varepsilon r+\xi)}F(\gamma-ig,2\gamma+1,-2i\varepsilon r)
    \right\}.
\end{eqnarray}
Here $\gamma=\sqrt{M^2-g^2}$, and the  real value $\xi$ satisfies the
equation $e^{-2i\xi}={(\gamma-ig)}/{M}$.

The asymptotics of (\ref{a}) are

%\begin{widetext}
\begin{eqnarray}\label{as}
&& \Xi_m=\begin{array}{c}
      {\rm Im }\\
      {\rm Re} \\
    \end{array}\Bigg\{\frac{\Gamma (2 \gamma +1) i^{-\gamma +i g}
   \sqrt{\gamma +i g} e^{i r \epsilon } (2r \epsilon
   )^{-1+i g}}{\Gamma (i g+\gamma +1)}\Bigg\}, \nonumber\\
   &&\phantom{\Xi_m=(2\varepsilon r)^{\gamma-1}\left(\begin{array}{c}
      { \sin\xi}\\
      { \cos\xi} \\
    \end{array}\right)
    , ~~~~}\mbox{at}~~ \epsilon r\to \infty\\
   &&  \Xi_m=(2\varepsilon r)^{\gamma-1}\left(\begin{array}{c}
      { \sin\xi}\\
      { \cos\xi} \\
    \end{array}\right)
    ,~~~~  \mbox{at}~~ \epsilon r\to 0\nonumber
\end{eqnarray}
%\end{widetext}

The plane wave with the fixed translational momentum $\bf k$ in absence of the Coulomb potential can be expanded into the radial the
waves
as
\begin{eqnarray}\label{bes}
\Psi_{\bf k}^{0}=\frac{1}{\sqrt{2\ k}}\sum_mi^me^{im(\varphi-\varphi_{\bf k})}\Bigg(\begin{array}{c}
      {J_m(kr) }\\
      {ie^{i\varphi}J_{m+1}(kr)} \\
    \end{array}\Bigg)
\end{eqnarray}
Here $\varphi_{\bf k}$  and $\varphi$ are the polar angles of  ${\bf k}$ and ${\bf r}$, correspondingly. The solution
(\ref{as}) is normalized to the unite flux in the  plane wave. At $r\to\infty$

\begin{eqnarray}\label{besas}
\Psi_{\bf k}^0= \frac{1}{\sqrt{\pi kr}}\sum_mi^me^{im(\varphi-\varphi_{\bf k})}\Bigg(\begin{array}{c}
       \cos \left(\frac{\pi
   m}{2}-kr+\frac{\pi }{4}\right) \\
      -i{e^{i\varphi} \sin \left(\frac{\pi
  m}{2}-kr+\frac{\pi }{4}\right)} \\
    \end{array}\Bigg)
\end{eqnarray}
To find the  cross-section of the process, one should relate the solutions (\ref{as}) and (\ref{besas}) at the infinity so that the coefficients at the divergent (or convergent) waves in each solutions coincide for the incoming (or outgoing) solutions.
The presence of the Coulomb potential at the large distance leads to a logarithmic phase change; this change should be
neglected,
because it has no effect on the flux.

The angular-momentum expansion of the true Coulomb wave function with  a given momentum $\bf k$ reads
\begin{equation}\label{psi}\Psi_{\bf k}^{d,c}=\sum_mc_m^{d,c}e^{im\varphi}\left(\begin{array}{cc}
                                                                              1 & 0 \\
                                                                              0 & e^{i\varphi}
                                                                            \end{array}\right)
\Xi_m,\end{equation}
where superscripts d and c refer to the wave functions obtained by  equating the coefficients at  the divergent $(\propto e^{ikr})$
or convergent $(\propto e^{-ikr})$ parts of the standing radial waves in $\Psi_{\bf k}$ in Eq.(\ref{besas}). Equating the
asymptotics
$$\sum_mc_m^{d,c}e^{im\varphi}\left(\begin{array}{cc}
                                                                              1 & 0 \\
                                                                              0 & e^{i\varphi}
                                                                            \end{array}\right)
\Xi_m^{d,c}=\Psi_{\bf k}^0$$
at $r\sim r_0$, where $r_0$ is some distance from the center larger than $1/\epsilon$ (latter on $r_0$ will disappear from the result),  we obtain

\begin{eqnarray}\label{cm}c_m^c=\frac{(1-i)  (-1)^m (2r_0 \epsilon )^{i g} e^{-im \varphi_{\bf k}+i
   \alpha_m }}{\sqrt{\pi } R_m},\nonumber\\
   \label{cd}c_m^d=-\frac{(1+i) (-1)^m(2r_0 \epsilon )^{-i g} e^{-i (\alpha_m -m \varphi_{\bf
   k})}}{\sqrt{\pi } R_m},\\\nonumber
   R_me^{i\alpha_m}=\frac{\Gamma (2 \gamma +1) i^{-\gamma +i g}
   \sqrt{\gamma +i g}  }{\Gamma (i g+\gamma +1)}.
   \end{eqnarray}
     We utilized the facts that the divergent wave in our case  belongs to the ${-\bf K}$ point and according to symmetry relations, the corresponding wave function can be found by  transformation $\varphi_{\bf
   k}\to -\varphi_{\bf
   k}$.
   \section*{Intervalley scattering}

Consider now the probability of the intervalley transitions caused by $\hat{U}$  taking into account the
finite-state Coulomb interaction. The amplitude of the intervalley transitions is $\langle\Psi_{\bf k'}^d|\hat{U}|\Psi_{\bf
k}^c\rangle$, where $\Psi_{\bf k}^{d,c}$ are determined by Eq. (\ref{psi}).  Note that only the terms with $m=m'$ and $m=m'\pm 1$ in $\Psi_{\bf k}^{d,c}$ have not been vanished after the angular integration. Subsequent radial integration leaves the   only
divergent term $m=m'=0$. As a result,
   \begin{eqnarray}\label{result}\nonumber&&|A_{\bf K, k;-K,k'}|^2=\frac{1}{{\cal A}^2}\frac{16}{\pi^2}\Bigg|\frac{\Gamma(\gamma+1+ig)}{\Gamma(2\gamma+1)}\Bigg|^4(2a\epsilon)^{2\sqrt{1-4g^2}-2}
\\&&\times U_0^2\Bigg(1-\sin
   ^2(2 \xi_0) \cos
   ^2\left(\frac{\varphi_{\bf k}+\varphi_{\bf
   k'}}{2}\right) \Bigg)\end{eqnarray}

The intervalley relaxation time is found by summation of all transitions in a box with $n_i{\cal A}$ impurities in it.

The    averaging over $\varphi_{\bf k'}$ gives the intervalley relaxation time

\begin{eqnarray}\label{result}\nonumber&&\frac{1}{\tau_v}=\frac{16n_i\epsilon}{\pi^2\hbar^3s^2}
\Bigg|\frac{\Gamma(\gamma+1+ig)}{\Gamma(2\gamma+1)}\Bigg|^4F_Z
 U_0^2\Bigg(1-2g^2  \Bigg),\\&& F_Z=(2a\epsilon/s\hbar)^{2\sqrt{1-4g^2}-2}.\end{eqnarray}
We revived $\hbar$ in this final formula. Eq. (\ref{result}) is not exact. Its accuracy is logarithmic:  $\log \langle\Psi_{\bf k'}^d|\hat{U}|\Psi_{\bf
k}^c\rangle|^2={2\sqrt{1-4g^2}-2}\log(1/2a\epsilon)+o(\log(1/2a\epsilon))$. At $2a\epsilon)\sim 1$ or $g\ll 1$  this result transforms
to the Born approximation.

Our consideration is limited by the case of $g^2<1/4$. If, additionally,  $g^2\ll 1/4$, the expansion
yields $F_Z=(2a\epsilon)^{-4g^2}$. One may express the intervalley scattering rate in the Coulomb field $1/\tau_v$ via the perturbative one  $1/\tau_{v, 0}$ as $1/\tau_v=F_Z/\tau_{v, 0}$.

While the Born intervalley relaxation rate drops when the energy goes down, the factor $F_Z=(2a\epsilon)^{2\sqrt{1-4g^2}-2}$ plays the role of the enhancement factor due
to
the Coulomb interaction. At a small $g$ this enhancement is weak, but essential.

It should be emphasized that the enhancement is independent of the sign of $g$, {\it e.g.} in the case of a repulsive
potential the enhancement also takes place! The source of this phenomenon lies in the conversion of the electrons to hole states
under the barrier surrounding the repulsive impurity with subsequent attraction to the impurity core.

At $|g|>\sqrt{3}/4$ the growth of the intervalley scattering rate with the energy changes to the drop.
Let us make some estimations. The constant $g=2e^2/\hbar s(\kappa+1)$ for graphene on the substrate with the dielectric constant $\kappa=14$ has the value $0.3$. In this case,  the power $(2\sqrt{1-4g^2}-2)=-0.422$. For electron energy  value $\epsilon= 1~$mV, $F_Z=20.4$.  For $U_0=1$~eV~$S_0$ and  impurity concentration $n=10^{12}$ cm${}^{-2}$, we have $\tau_{v}=1.437\times 10^{-9}~$s.

\section*{Discussion and conclusions}

We have found the intervalley scattering rate with and without Coulomb interaction. While the short-range interaction gives rise to the intervalley  amplitude that is independent of the electron energy, the long-range Coulomb interaction also contributes to the process via the Sommefield prefactor power-like depending on the electron energy. Our consideration is limited by the weak enough electrostatic interaction constant $|g|<1/2$. Strictly speaking, this is not the case of the free-suspended graphene, but in most cases of the graphene on the semiconductor substrate this condition is fulfilled. The case of $|g|>1/2$ can not to be considered in a single-electron approximation due to the falling-down-to-origin phenomenon. We have found that  the Sommefield prefactor has an attractive character for any the interaction sign. This essentially differs the graphene case from a gap-band semiconductor.

The intervalley relaxation time found here controls the process of the valley population relaxation. It should be emphasized that the analogy between the pseudospin and spin has a wider meaning than  the valley population. For example,  coherent electron states in  different valleys can be constructed by optical orientation processes where optical transitions  produce electrons in both valleys simultaneously and, hence, coherently. Generally speaking, these mixed states decay in other way than  the established equilibration  of the valley population.  The relaxation of these coherent states is similar to the transversal spin relaxation. The consideration of this relaxation goes beyond the scope of the present paper.

\section*{Acknowledgements}
The work was supported by the RFBR grants 13-0212148 and 14-02-00593.

\end{document}